\documentclass[a4paper,12pt]{article}
\usepackage[dvips]{graphicx}
\usepackage{epsfig}
\usepackage{graphicx}
\usepackage{color}
\usepackage{amssymb}
\usepackage{amsfonts}
\usepackage{amsmath}
\usepackage{xcolor}
\usepackage{hyperref}
\usepackage{comment}

\begin{document}
\title{Useful formulas for non-magnetized electron cooling}
\author{S. Seletskiy\footnote{seletskiy@bnl.gov}, A. Fedotov}
\date{\today}

\maketitle

\begin{abstract}
Recent success of Low Energy RHIC Electron Cooler (LEReC) opened a road for development of high energy electron coolers based on non-magnetized electron bunches accelerated by RF cavities. Electrons in such coolers can have velocity distribution with various unequal horizontal, vertical and longitudinal velocity spreads. In this paper we revisit a formula of friction force in non-magnetized cooling and derive a number of useful expressions for different limiting cases.
\end{abstract}

\section{Introduction}
Electron Cooling (EC) \cite{Budker, BudkerDemo} is a technique that allows increasing a 6-D phase space density of stored hadron beams. 

In EC a beam of ``cold'' electrons co-propagate with a hadron beam with the same average velocity in a straight section of the storage ring, called a cooling section (CS).  A hadron interacts with electrons in a CS via Coulomb force, which introduces dynamical friction \cite{DynamicalFriction} acting on each hadron. After every passage through the CS the electrons are either dumped or returned to the electron gun for charge recovery, thus, on each turn the ions interact with fresh electrons. Over many revolutions in the accelerator the average friction reduces both the transverse and the longitudinal momentum spread of the ion bunch.

A typical nonrelativistic (with $\gamma$-factor $\lesssim 1$) electron cooler utilizes a magnetized DC electron beam, that is both the cathode and the CS are immersed in the solenoidal field. A recent success of Low Energy RHIC Electron Cooler \cite{LEReCFedotov, LEReCSeletskiy1, LEReCGu, LEReCKayran, LEReCSeletskiy2}, the first RF-based non-magnetized electron cooler, demonstrated that operational EC does not require e-beam magnetization. While LEReC operated at $\gamma=4.1$ and $\gamma=4.6$, LEReC approach  becomes especially attractive for high energy electron coolers because it significantly simplifies the engineering aspects of a cooler design. For example, a pre-cooler \cite{EICPrecooler} for Electron Ion Collider \cite{EIC} (which will be built at BNL) will utilize non-magnetized RF-accelerated bunches of electrons.

In this note we revisit the general expression for the cooling force in a non-magnetized EC and derive both useful analytical formulas for several limiting cases and expressions convenient for the fast numerical integration. While some of the formulas presented in this paper are well known, others, to the best of our knowledge, have never being published before.

\section{Electron bunches with Maxwell distribution}

The friction force acting on an ion co-traveling with  an electron bunch with velocity distribution $f(v_e)$ is given by \cite{ChandraFriction, Derben78}:

\begin{equation}\label{eq1Md}
    \vec{F}=-\frac{4 \pi n_e e^4 Z^2}{m_e} \int{L_C \frac{\vec{v}-\vec{v}_e}{\left| \vec{v}-\vec{v}_e \right|^3} f(v_e) d^3v_e}
\end{equation}
Here, $n_e$ is the electron bunch density in the beam frame,  $e$ is the electron charge, $Z \cdot e$ is the ion charge, $m_e$ is the mass of the electron, $\vec{v}$ and $\vec{v}_e$ are ion and electron velocities in the beam frame and $L_C$ is the Coulomb logarithm, which has a weak dependence on $v_e$ and can be moved from under the integral.

If we assume a Maxwell-Boltzmann distribution of velocities in the electron bunch then Eq. (\ref{eq1Md}) can be simplified to 1D integrals for each component of the friction force \cite{Binney}. Such formulas (named Binney's formulas) are very convenient for numerical integration. 

In  this paper we assume that there is no offset between average velocities of electrons and ions. Derivation of Binney's formulas for v-distribution with a constant offset between beams' velocities  along with considerations of relevant physics can be found in \cite{ourBinney}.

The electron bunch v-distribution is given by:

\begin{equation}\label{eq2Md}
f(v_e)=\frac{1}{(2\pi)^{3/2} \Delta_x \Delta_y \Delta_z} \exp \left( 
-\frac{v_{ex}^2}{2\Delta_x^2}-\frac{v_{ey}^2}{2\Delta_y^2}-\frac{v_{ez}^2}{2\Delta_z^2}
                                                                               \right)
\end{equation}

Let us introduce an effective potential in a velocity-space:

\begin{equation} \label{eq3Md}
U=C_0 \int \frac{f(v_e)}{\left| \vec{v}-\vec{v}_e \right|} d^3 v_e
\end{equation}
such that 
\begin{equation} \label{eq3aMd}
F_{x,y,z}=\partial U / \partial v_{x,y,z}
\end{equation}
Here, $C_0=\frac{4 \pi n_e e^4 Z^2 L_C}{m_e}$.

Noticing that $1/\left| \vec{v}-\vec{v}_e \right|$ can be represented as:

\begin{equation} \label{eq4Md}
\begin{aligned}
\frac{1}{\sqrt{(v_x-v_{ex})^2+(v_y-v_{ey})^2+(v_z-v_{ez})^2}} = \\
\frac{2}{\sqrt{\pi}} \int_0^\infty \exp\left[-p^2 \left(
 (v_x-v_{ex})^2+(v_y-v_{ey})^2+(v_z-v_{ez})^2
                                                  \right) \right]dp 
\end{aligned}
\end{equation}
we get from Eqs. (\ref{eq2Md})-(\ref{eq4Md}):

\begin{equation} \label{eq5Md}
\begin{aligned}
U =& \frac{C_0}{\sqrt{2}\pi^2 \Delta_x \Delta_y \Delta_z} \int_0^\infty \left[ 
       \int_{-\infty}^\infty e^{ -p^2 (v_x-v_{ex})^2 - \frac{v_{ex}^2}{2\Delta_x^2} } dv_{ex} \right. \\
     & \left. \int_{-\infty}^\infty e^{ -p^2 (v_y-v_{ey})^2 - \frac{v_{ey}^2}{2\Delta_y^2} } dv_{ey}
               \int_{-\infty}^\infty e^{ -p^2 (v_z-v_{ez})^2 - \frac{v_{ez}^2}{2\Delta_z^2} } dv_{ez} 
       \right]dp 
\end{aligned}
\end{equation}

Since:
\begin{equation} \label{eq6Md}
\int_{-\infty}^\infty e^{ -p^2 (v_x-v_{ex})^2 - \frac{v_{ex}^2}{2\Delta_x^2} } dv_{ex}=
\sqrt{2\pi} \Delta_x \frac{\exp \left( -\frac{p^2 v_x^2}{1+2p^2\Delta_x^2}\right)}{\sqrt{1+2p^2\Delta_x^2}}
\end{equation}
we get for the effective potential:
\begin{equation} \label{eq7Md}
U=\frac{2C_0}{\sqrt{\pi}} \int_0^\infty 
\frac{\exp \left( -\frac{p^2 v_x^2}{1+2p^2\Delta_x^2}
                      -\frac{p^2 v_y^2}{1+2p^2\Delta_y^2}
                      -\frac{p^2 v_z^2}{1+2p^2\Delta_z^2} \right)}
      {\sqrt{(1+2p^2\Delta_x^2)(1+2p^2\Delta_y^2)(1+2p^2\Delta_z^2)}}
                                     dp
\end{equation}

From Eqs. (\ref{eq3aMd}) and (\ref{eq7Md}) we obtain Binney's formulas. For example, for $F_x$:

\begin{equation} \label{eq8Md}
F_x=-\frac{4C_0}{\sqrt{\pi}} \int_0^\infty \frac{p^2 v_x}{1+2p^2\Delta_x^2}
\frac{\exp \left( -\frac{p^2 v_x^2}{1+2p^2\Delta_x^2}
                      -\frac{p^2 v_y^2}{1+2p^2\Delta_y^2}
                      -\frac{p^2 v_z^2}{1+2p^2\Delta_z^2} \right)}
      {\sqrt{(1+2p^2\Delta_x^2)(1+2p^2\Delta_y^2)(1+2p^2\Delta_z^2)}}
                                     dp
\end{equation}

A more elegant form of Binney's formulas is obtained by substituting $p^2=1/(2q\Delta_x^2)$ into expressions (\ref{eq8Md}):

\begin{equation}\label{eq9Md}
\left\{
\begin{array}{lcl}
F_x & = & -C v_x \int_0^\infty \frac{\exp \left[ -\frac{v_x^2}{2 \Delta_x^2 (1+q)} -\frac{v_y^2}{2 (\Delta_x^2 q+ \Delta_y^2)}- \frac{v_z^2}{2(\Delta_x^2 q +\Delta_z^2)}\right]}{\Delta_x^3(1+q)^{3/2} (q +\Delta_y^2/\Delta_x^2)^{1/2}(q +\Delta_z^2/\Delta_x^2)^{1/2}} dq\\
F_y & = & -C v_y \int_0^\infty \frac{\exp \left[ -\frac{v_x^2}{2 \Delta_x^2 (1+q)} -\frac{v_y^2}{2 (\Delta_x^2 q+ \Delta_y^2)}- \frac{v_z^2}{2(\Delta_x^2 q +\Delta_z^2)}\right]}{\Delta_x^3(q+\Delta_y^2/\Delta_x^2)^{3/2} (q +1)^{1/2}(q +\Delta_z^2/\Delta_x^2)^{1/2}} dq\\
F_z & = & -C v_z \int_0^\infty \frac{\exp \left[ -\frac{v_x^2}{2 \Delta_x^2 (1+q)} -\frac{v_y^2}{2 (\Delta_x^2 q+ \Delta_y^2)}- \frac{v_z^2}{2(\Delta_x^2 q +\Delta_z^2)}\right]}{\Delta_x^3(q+\Delta_z^2/\Delta_x^2)^{3/2} (q +1)^{1/2}(q +\Delta_y^2/\Delta_x^2)^{1/2}} dq\\
\end{array}
\right.
\end{equation}
where $C=2\sqrt{2\pi}n_e r_e^2 m_e c^4 Z^2 L_C$.

For a special case of $\Delta_x=\Delta_y \equiv \Delta_t$ Eq. (\ref{eq9Md}) becomes:

\begin{equation}\label{eq10Md}
\left\{
\begin{array}{rcl}
F_{x,y} & = & -C v_{x,y} \int_0^\infty g_t(q) dq\\
F_z & = & -C v_z \int_0^\infty g_z(q) dq\\
g_t(q) & = & \frac{1}{\Delta_t^2(1+q)^2 \sqrt{\Delta_t^2 q +\Delta_z^2}} \exp \left[ -\frac{v_x^2+v_y^2}{2 \Delta_t^2 (1+q)} - \frac{v_z^2}{2(\Delta_t^2 q +\Delta_z^2)}\right]\\
g_z(q) & = & \frac{1}{(1+q) (\Delta_t^2 q +\Delta_z^2)^{3/2}} \exp \left[ -\frac{v_x^2+v_y^2}{2 \Delta_t^2 (1+q)} - \frac{v_z^2}{2(\Delta_t^2 q +\Delta_z^2)}\right]\\
\end{array}
\right.
\end{equation}

For the case of an isotropic distribution ($\Delta_x=\Delta_y=\Delta_z \equiv \Delta$) Eqs. (\ref{eq9Md}, \ref{eq10Md}) reduce to Chandrasekhar's classical result: 

\begin{equation}\label{eq11Md}
\left\{
\begin{array}{rcl}
F_{x,y,z} & = & -\frac{C}{\Delta^3 a^3}  v_{x,y,z}\left[-2ae^{-a^2/2}+\sqrt{2\pi}\mathrm{Erf}\left(\frac{a}{\sqrt{2}}\right)\right]\\
a & = & \sqrt{v_x^2+v_y^2+v_z^2}/\Delta \\
\end{array}
\right.
\end{equation}
where the error function $\mathrm{Erf(z)}\equiv \frac{2}{\sqrt{\pi}}\int_0^z e^{-t^2}dt$.

Expressions for friction force components given by Eq. (\ref{eq9Md}) are useful for designing a bunched electron cooler, since for such a cooler a velocity distribution of electrons might have non-equal horizontal, vertical and longitudinal spreads.

\section{Limiting cases for hadrons with small velocities}

\subsection{e-bunch with arbitrary Maxwell distribution}

\begin{figure}[!htb]
  \centering
  \includegraphics[width=0.7\columnwidth]{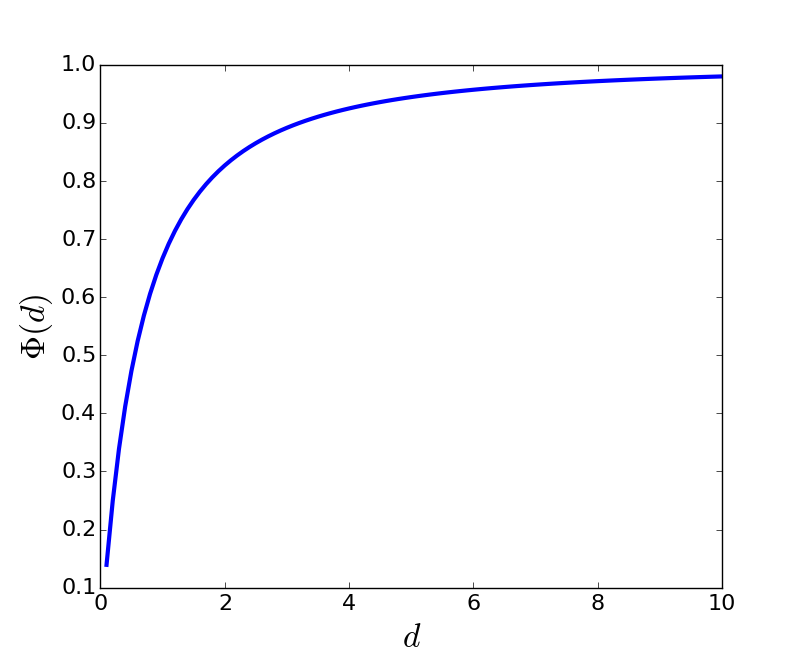}
  \caption{Function $\Phi(d)$.}
   \label{Phi}
\end{figure}

The integrals \ref{eq10Md} can be taken analytically in the approximation of a ``linear'' friction force, i.e. when $v_{x,y,z} \ll \Delta_{x,y,z}$. The resulting expression is given by Eq. (\ref{eq1Lc}).

\begin{equation}\label{eq1Lc}
\begin{array}{lcl}
F_{x,y}  &=& -C \frac{v_{x,y}}{\Delta_\perp^2 \Delta_z} \Phi(\Delta_z/\Delta_\perp) \\
F_z  &=& -2C \frac{v_z}{\Delta_\perp^2 \Delta_z} (1-\Phi(\Delta_z/\Delta_\perp)) \\
\Phi(d) & = & \left\{ \begin{array}{l} 
\frac{d}{1-d^2} \left( \frac{\arccos(d)}{\sqrt{1-d^2}} -d \right), \ d<1 \\
2/3, \ d=1 \\
\frac{d}{d^2-1} \left( \frac{\log(d-\sqrt{d^2-1})}{\sqrt{d^2-1}} +d \right), \ d>1 \\
\end{array}\right.\\
\end{array}
\end{equation}
where $d=\Delta_z/\Delta_\perp$. Note that $\Phi$ is a continuous function (see Fig \ref{Phi}). By expanding $\Phi(d)$ both for $d>1$ and for $d<1$ in Taylor series around $d=1$ one can easily check that
\[
\lim \limits_{d \to 1} \Phi(d<1) = \lim \limits_{d \to 1} \Phi(d>1)=\frac{2}{3}
\] 

Formulas (\ref{eq1Lc}) give the friction force experienced by the small velocity ions in a general case (no assumptions about the relation between $\Delta_\perp$ and $\Delta_z$ is made) of Maxwell distribution of the electron bunch velocities ($f(v_e)$). It is not too surprizing that our formulas remind of the Ogino-Ruggiero formulas derived for the case of the elliptic uniform distribution $f(v_e)$ (see equations (26a-26b) in \cite{Ogino}).

Equation (\ref{eq1Lc}) is useful for quick estimates of the cooling force in case of anisotropic velocity distribution in the e-bunch. 

Next, we will compare the derived equations to the numeric integration of Binney's formulas for several distributions with various $\Delta_\perp$ and $\Delta_z$. For this test we will use the typical LEReC parameters for the 2 MeV settings.

\begin{figure}[!htb]
  \centering
  \includegraphics[width=1\columnwidth]{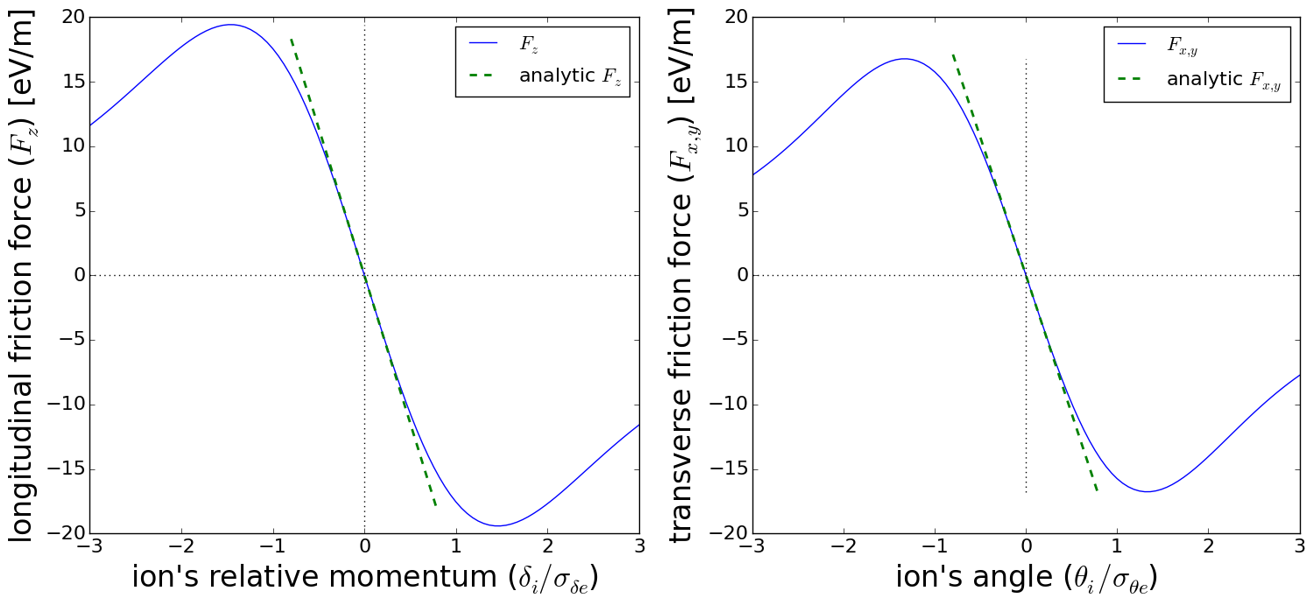}
  \caption{Comparison of friction force obtained by numerical integration of Binney's formulas (\ref{eq10Md}) to the linear approximation (\ref{eq1Lc}) for the case of $\Delta_z/\Delta_\perp=0.68$.}
   \label{LEReC_d07}
\end{figure}

\begin{figure}[!htb]
  \centering
  \includegraphics[width=1\columnwidth]{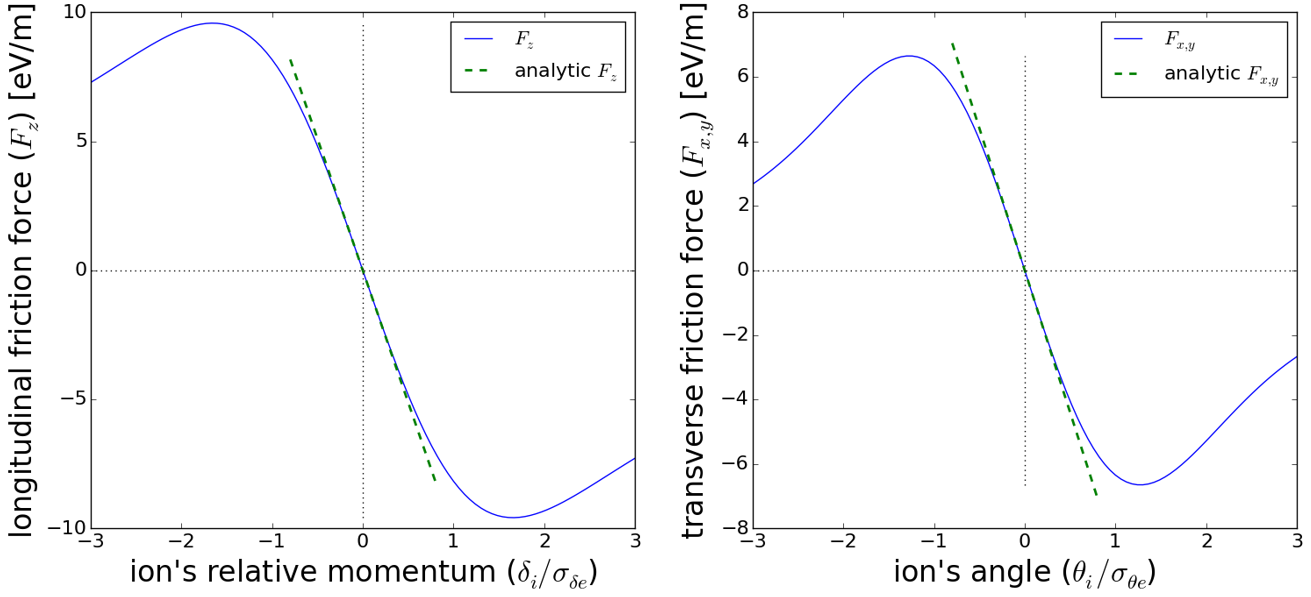}
  \caption{Comparison of friction force obtained by numerical integration of Binney's formulas (\ref{eq10Md}) to the linear approximation (\ref{eq1Lc}) for the case of $\Delta_z/\Delta_\perp=0.34$.}
   \label{LEReC_d03}
\end{figure}

Figure \ref{LEReC_d07} shows the results of the comparison for the case of $\sigma_{\theta e}=150$ $\mu$rad and $\sigma_{\delta e}=5\cdot10^{-4}$, which corresponds to $d \approx 0.68$. Similarly Fig. \ref{LEReC_d03} shows the results of the comparison for the case of $\sigma_{\theta e}=300$ $\mu$rad and $\sigma_{\delta e}=5\cdot10^{-4}$, which corresponds to $d \approx 0.34$. As one can see Eq. (\ref{eq1Lc}) give the perfect approximation of the linear part of the friction force. 

Finally, we will consider two limiting cases: $\Delta_z=\Delta_\perp$ and  $\Delta_z \ll \Delta_\perp$.

\subsection{e-bunch with isotropic Maxwell distribution}
For $\Delta_z=\Delta_\perp \equiv \Delta$ we have $d=1$. Hence, we get:
\begin{equation}\label{eq2Lc}
F_{x,y}=F_z  =-\frac{4\sqrt{2\pi}}{3}\frac{n_e r_e^2 m_e c^4 Z^2 L_C}{\Delta^3}v_{x,y,z}
\end{equation}
which is a well known result (see, for instance, equation (1.12) in \cite{derbDisser}).

\subsection{e-bunch with strongly anisotropic Maxwell distribution}
For $\Delta_z \ll \Delta_\perp$ we must substitute $d \to 0$ into (\ref{eq1Lc}). We have $\Phi(d\to0) \to d \cdot \pi/2$. Therefore, for the components of the friction force we immediately get:

\begin{equation}\label{eq3Lc}
\begin{array}{rcl}
F_{x,y} &=&-\pi\sqrt{2\pi} \frac{n_e r_e^2 m_e c^4 Z^2 L_C}{\Delta_\perp^3}v_{x,y}\\
F_z  &=&-4\sqrt{2\pi} \frac{n_e r_e^2 m_e c^4 Z^2 L_C}{\Delta_\perp^2 \Delta_z}v_z\\
\end{array}
\end{equation}
The derived asymptotic expressions are also well-known (see formulas (1.49)-(1.50) in \cite{derbDisser}).

\section{Cooling rate}

Let us write down the cooling rate for the case of $\Delta_x=\Delta_y$. 

The cooling rate ($\lambda$) in the laboratory frame can be obtained from:

\begin{equation}\label{eq1Cr}
\lambda=\frac{F \eta}{\gamma m_i v}
\end{equation}
where the duty factor $\eta=L_{CS}/(2\pi R)$, $R$ is the storage ring radius, $L_{CS}$ is the length of the cooling section, $m_i=A_i m_p$, $m_p$ is the proton mass and $A_i$ is the ion mass number.

Notice that the e-bunch density in the beam frame $n_e=\frac{1}{\gamma}\frac{N_e}{(2\pi)^{3/2} \sigma_x \sigma_y \sigma_z}$, where $\sigma_{x,y,z}$ are the horizontal, the vertical and the longitudinal e-bunch sizes in the laboratory frame. Also, $\Delta_z=\beta c \sigma_\delta$, and $\Delta_\perp=\gamma \beta c \sigma_{\theta}$, where $\sigma_\theta \equiv \sigma_{\theta x}=\sigma_{\theta y}$, Then, substituting Eq. (\ref{eq1Lc}) into Eq. (\ref{eq1Cr}),  we get for the transverse and the longitudinal cooling rates defined strictly through the laboratory frame beam parameters:

\begin{equation}\label{eq2Cr}
\begin{array}{lcl}
\lambda_{x,y}  &=& -\frac{N_e r_e^2 m_e c Z^2 L_C}{\pi \gamma^4 \beta^3 A_i m_p \sigma_x \sigma_y \sigma_z} \frac{\eta}{\sigma_\theta^2 \sigma_\delta} \Phi\left( \frac{\sigma_\delta}{\gamma \sigma_{\theta}} \right) \\
\lambda_{z}  &=& -\frac{2 N_e r_e^2 m_e c Z^2 L_C}{\pi \gamma^4 \beta^3 A_i m_p \sigma_x \sigma_y \sigma_z} \frac{\eta}{\sigma_\theta^2 \sigma_\delta} \left( 1-\Phi\left( \frac{\sigma_\delta}{\gamma \sigma_{\theta}} \right) \right) \\
\Phi(d) & = & \left\{ \begin{array}{l} 
\frac{d}{1-d^2} \left( \frac{\arccos(d)}{\sqrt{1-d^2}} -d \right), \ d<1 \\
2/3, \ d=1 \\
\frac{d}{d^2-1} \left( \frac{\log(d-\sqrt{d^2-1})}{\sqrt{d^2-1}} +d \right), \ d>1 \\
\end{array}\right.\\
\end{array}
\end{equation}
Please notice that using the laboratory frame parameters: $d=\sigma_\delta/(\gamma \sigma_{\theta})$.

\section{Conclusion}
In this paper we derived several expressions useful in design of the non-magnetized bunched electron coolers.

Equations (\ref{eq9Md})-(\ref{eq11Md}) give the friction force for an electron bunch with the Maxwell-Boltzmann velocity distribution in a form of one-dimensional integrals. 

Equation (\ref{eq1Lc}) gives an analytic expression for the friction force acting on  ions with velocities smaller than the electron bunch velocity spreads in a general case of unequal longitudinal and transverse velocity spreads of electrons.

Equation (\ref{eq2Cr}) gives an explicit analytic formula for the cooling rate expressed in the laboratory frame parameters of the electron and ion beams.

\end{document}